\documentstyle[prd,aps,epsfig,floats]{revtex}
 
\newcommand{\be}{\begin{equation}}
\newcommand{\ee}{\end{equation}}
\newcommand{\ben}{\begin{eqnarray}}
\newcommand{\een}{\end{eqnarray}}
\newcommand{\bc}{\begin{center}}
\newcommand{\ec}{\end{center}}
\newcommand{\tn}{\tablenote}

\newcommand{\D}{\Delta}

\renewcommand{\phi}{\varphi}

\begin{document}
\draft
\twocolumn[
\hsize\textwidth\columnwidth\hsize\csname
@twocolumnfalse\endcsname\widetext

\title{Microlensing by natural wormholes: theory and simulations}

\author{Margarita Safonova
\cite{byline1}}
\address{Department of Physics and Astrophysics, University of Delhi,
New Delhi--7, India}
\author{Diego F. Torres\cite{byline2} and
Gustavo E. Romero\cite{byline3}}
\address{Instituto Argentino de Radioastronom\'{\i}a, C.C.5,
1894 Villa Elisa, Buenos Aires, Argentina}

\maketitle
\begin{abstract}
We provide an in depth study of the theoretical peculiarities that
arise in effective negative mass lensing, both for the case of a
point mass lens and source, and for extended source situations. We
describe novel observational signatures arising in the case of a
source lensed by a negative mass. We show that a negative mass
lens produces total or partial eclipse of the source in the umbra
region and also show that the usual Shapiro time delay is replaced
with an equivalent time gain. We describe these features both
theoretically, as well as through numerical simulations. We
provide negative mass microlensing simulations for various
intensity profiles and discuss the differences between them. The
light curves for microlensing events are presented and contrasted
with those due to lensing produced by normal matter. Presence or
absence of these features in the observed microlensing events can
shed light on the existence of natural wormholes in the Universe.

\end{abstract}


\vskip2pc
]\renewcommand{\thefootnote}{\arabic{footnote}}
\narrowtext

\section{Introduction}

Wormhole solutions to the Einstein field equations have been
extensively studied in the last decade (see Refs. \cite{motho,wh}
and references cited therein, as well as the book by Visser
\cite{VISSER-BOOK}). Wormholes basically represent {\em bridges}
between otherwise separated regions of the space-time (see
Fig.~1) and need a special kind of matter in order to exist.
This matter, known as exotic, violates the energy conditions (EC),
particularly the null (or averaged null) one
\cite{VISSER-BOOK,h-v,h-v2}.

To specify what we are referring to when talking about the energy
conditions, we shall provide their point-wise form. Apart from the
null (NEC), they are the weak (WEC), the strong (SEC), and the
dominant (DEC) energy conditions. For a Friedman-Robertson-Walker
space-time and a diagonal stress-energy tensor
$T_{\mu\nu}=(\rho,-p,-p,-p)$ with $\rho$ the energy density and
$p$ the pressure of the fluid, they read:
\begin{eqnarray}
\hbox{NEC} & \iff &  \quad (\rho + p \geq 0 ), \nonumber \\
\hbox{WEC} & \iff & \quad (\rho \geq 0 ) \hbox{ and } (\rho + p
\geq 0),  \nonumber \\ \hbox{SEC} & \iff & \quad (\rho + 3 p \geq
0 ) \hbox{ and } (\rho + p \geq 0), \nonumber \\ \hbox{DEC} & \iff
& \quad (\rho \geq 0 ) \hbox{ and } (\rho \pm p \geq 0).
\end{eqnarray}
The EC are, then, linear relationships between the energy density
and the pressure of the matter generating the space-time
curvature. We can immediately see why the possible violations of
the EC are so polemical. If NEC is violated, then WEC is also
violated. Negative energy densities---and so negative masses---are
thus physically admitted. Nevertheless, it is important to keep in
mind that the EC of classical General Relativity are only
conjectures. They are widely used to prove theorems concerning
singularities and black hole thermodynamics, such as the area
increase theorem, the topological censorship theorem, and the
singularity theorem of stellar collapse \cite{VISSER-BOOK}.
However, all EC lack a rigorous proof and, indeed, several
situations in which the EC are violated are known; perhaps the
most quoted being the Casimir effect, see Refs.
\cite{wh,last-visser}. Typically, observed violations are produced
by small quantum systems, resulting of the order of $\hbar$. It is
currently far from clear whether there could be macroscopic
quantities of such an exotic, e.g. WEC-violating, matter. If it
does exist, macroscopic negative masses could be part of the
ontology of the universe.

In fact, the possible existence of negative gravitational masses
is being investigated at least since the end of the nineteenth
century \cite{JAMMER}. The empirical absence of negative masses in
the Earth neighborhood could be explained as the result of the
plausible assumption that, repelled by the positive masses
prevalent in our region of space, the negative ones have been
driven away to extragalactic distances. Bondi already remarked in
\cite{BONDI} that it is just an empirical fact that inertial
and gravitational masses are both positive quantities. Clearly, no
other way better than devising observational tests for deciding
the controversy on negative mass existence is available. For
instance, if natural wormholes actually exist in the universe
(e.g. if the original topology after the Big-Bang was multiply
connected), then there could be some observable electromagnetic
signatures that might lead to their identification.

\begin{figure}[t]
\begin{center}
\epsfxsize=3cm 
\epsffile{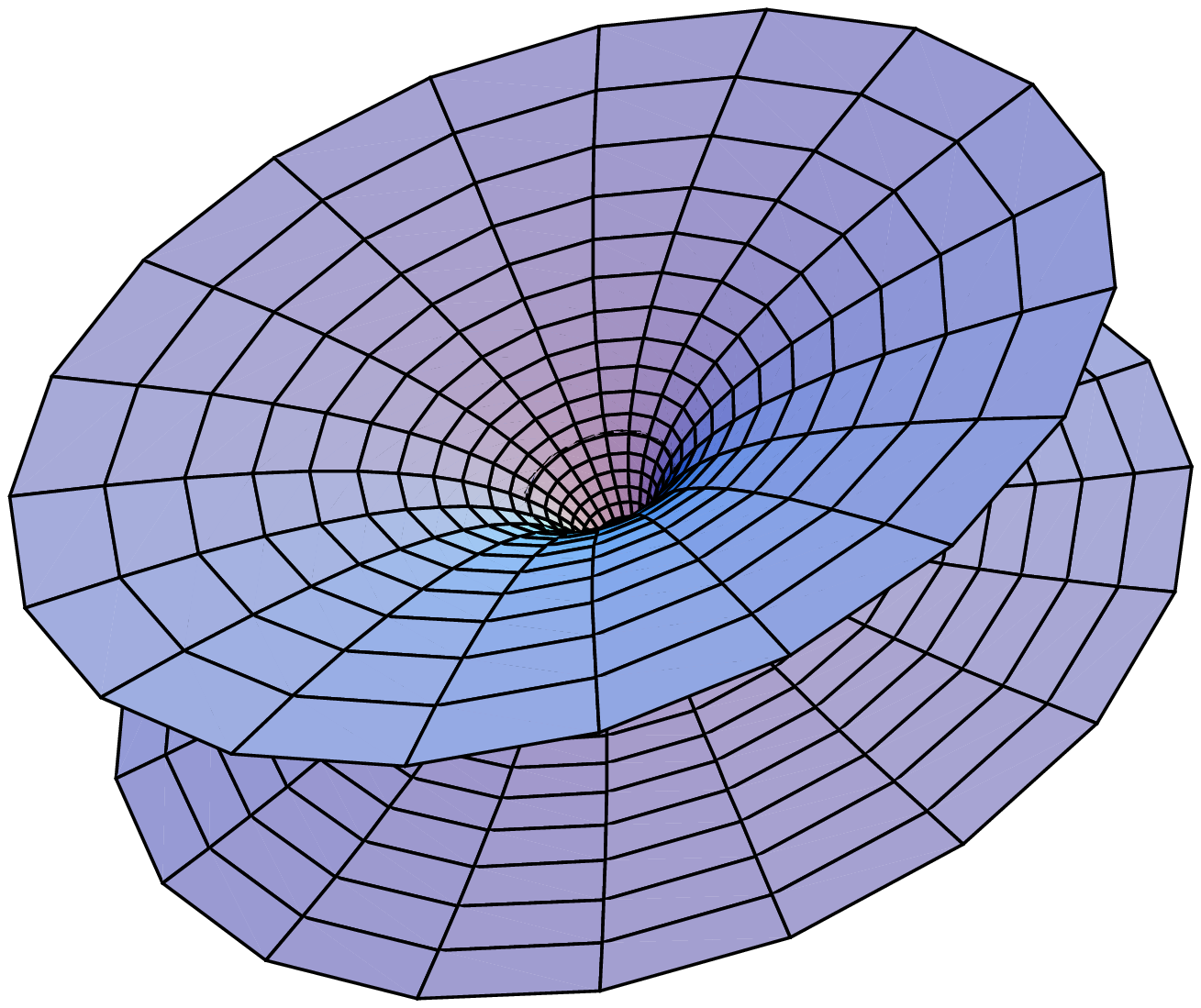} 
\hspace{.8cm}
\epsfxsize=3cm 
\epsffile{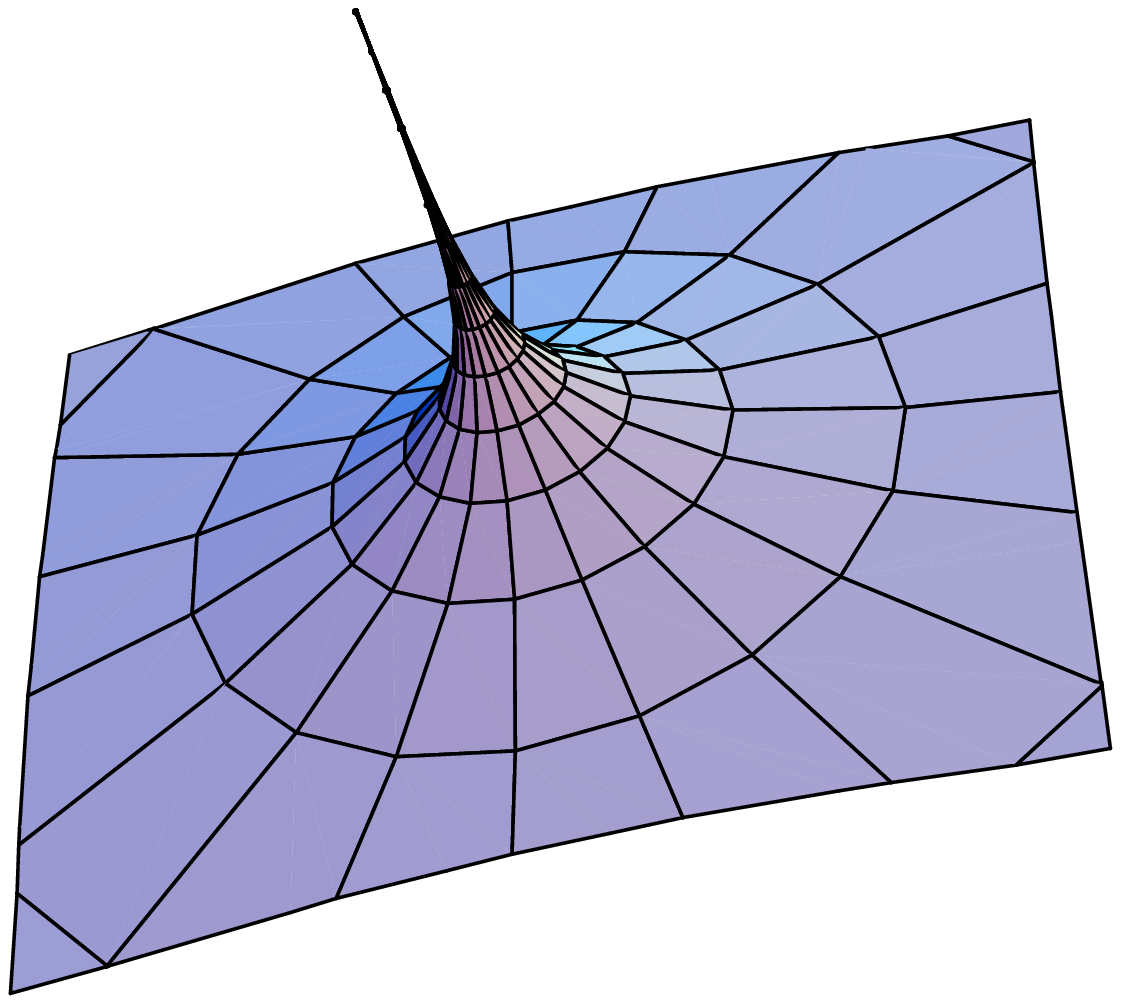}
\end{center} \vspace{.25cm} \caption{Left: Embedding diagram for a
wormhole. Two mouths, joined by a tunnel, can connect regions
otherwise very much separated (here the normal space should fold
as a sheet of paper, whereas the wormhole would be a tunnel from
one side of the sheet to the other). Right: Embedding diagram for
a black hole. The singularity here is represented as a pinch off
of the wormhole tunnel.}
\end{figure}

The idea that wormholes can act as gravitational lenses and induce
a microlensing signature on a background source was first
suggested by Kim and Sung \cite{swkim}. Unfortunately, their
geometry was of a perfect alignment of a source, both wormhole's
mouths and an observer, which is, on a common sense ground, quite
unlikely. They also considered both mouths to be of positive mass.
Cramer et al. \cite{cramer} carried out more detailed analysis of
a negative mass wormholes and considered the effect they can
produce on background point sources, at non-cosmological
distances. The generalization to a cosmological scenario was
carried out by Torres et al. \cite{diego-grbwh}, although lensing
of point sources was still used. As far as we are aware, the first
and only bound on the possible existence of negative masses,
imposed using astrophysical databases, was given by Torres et al.
\cite{diego-grbwh}. These authors showed that the effective
gravitational repulsion of light rays creates two bursts, which
are individually asymmetric under time reversal, although mirror
images of each other. Recently, Anchordoqui et al. \cite{doqui}
searched in existent gamma-ray bursts databases for signatures of
wormhole microlensing. Although they detected some interesting
candidates, no conclusive results could be obtained. Peculiarly
asymmetric gamma-ray bursts \cite{rom}, although highly uncommon,
might be probably explained by more conventional hypothesis, like
precessing fireballs (see, for instance, Ref. \cite{Zwart}).

The case of macrolensing was recently analyzed by us as well,
showing that if large concentrations of localized negative masses
do exist, we should be able to see distinctive effects upon a deep
background \cite{macro}. All in all, possible existence of
wormholes or of any kind of negative masses in the universe could
not yet be discarded. In order to confirm or to deny the existence
of such masses, we need to develop a strong theoretical framework
and to provide a clear test through observations.

In this work we revisit the microlensing by natural wormholes of
stellar and sub-stellar masses. We provide an in-depth study of
the theoretical peculiarities that arise in negative mass
microlensing, both for a point mass lens and source, and for
extended source situations. For the first time, we present
negative mass microlensing simulations, showing the resulting
shapes of the images, the intensity profiles, the time gain
function, the radial and tangential magnifications, and other
features. In this regard, this work extends and deepens previous
papers in several ways, and gives a complete description from
where to analyze, from a computational and quantitative point of
view, observational predictions, as the ones presented for
chromaticity in Ref. \cite{eiroa}.

This work is divided as follows. In Section II we introduce the
basic theoretical framework of gravitational lensing produced by a
generic negative mass, assumed here to be a wormhole. Two things
should be clear to the reader. The first is that the use of
wormholes is just to fix an interesting theoretical background.
Any strut of negative mass would produce the same effects. The
second is that we never consider light going through the wormhole
mouths, but being deflected in the neighborhood. We introduce the
effective refractive index, magnification results, the time gain
function, and other features, in several subsections of Section
II. Section III presents microlensing simulations, showing the
form and position of the produced images. In Section IV, we treat
the extended source case, also using a numerical code and taking
into account different source profiles. We finally give some
concluding remarks.

\section{Lensing by a point negative mass}

In this paper we consider lensing only by a point negative mass
lens, and thus we can use all the assumptions concurrent with the
treatment of the Schwarzschild lens:

\begin{itemize}
\item {\it Geometrical optics approximation}---the scale over which
the gravitational field changes is much larger than the wavelength of
the light being deflected.

\item {\it Small-angle approximation}---the total deflection angle is
small. The typical bending angles involved in gravitational
lensing of cosmological interest are less than $ < 1^{\prime}$;
therefore we describe the lens optics in the paraxial
approximation.

\item {\it Geometrically-thin lens approximation}---the maximum
deviation of the ray is small compared to the length scale on which the
gravitational field changes. Although the scattering takes place
continuously over the trajectory of the photon, the appreciable
bending occurs only within a distance of the order of the impact
parameter.

\end{itemize}

We begin the discussion of gravitational lensing by defining two
planes, the source and the lens plane. These planes,
described by Cartesian coordinate systems ($\xi_1,\xi_2$) and
($\eta_1,\eta_2$), respectively, pass through
the source and deflecting mass and are perpendicular to the
optical axis (the straight line extended from the source plane
through the deflecting mass to the observer). Since the components
of the image position and the source positions are much smaller in
comparison to the distances to lens and source planes, we can
write the coordinates in terms of the observed angles. Therefore,
the image coordinates can be written as $(\theta_1,\theta_2$) and
those of the source as $(\beta_1,\beta_2)$.

\subsection{Effective refractive index of the gravitational field
of a negative mass and the deflection angle}

The `Newtonian' potential of a negative point mass lens is given
by \be \Phi(\xi,z) = \frac{G |M|}{(b^2 +z^2)^{1/2}}\,\,, \ee where
$b$ is the impact parameter of the unperturbed light ray and $z$
is the distance along the unperturbed light ray from the point of
closest approach. We have used the term Newtonian in quotation
marks since it is, in principle, different from the usual one.
Here the potential is positive defined and approaching zero at
infinity \cite{VISSER-BOOK}. In view of the assumptions stated
above, we can describe light propagation close to the lens in a
locally Minkowskian space-time perturbed by the {\em positive}
gravitational potential of the lens to first post-Newtonian order.
In this weak field limit, we describe the metric of the negative
mass body in orthonormal coordinates $x^0=ct$, ${\bf x}=(x^i)$ by
\be ds^2 \approx \left(1 +  \frac{2 \Phi}{c^2}\right)c^2dt^2 -
\left(1 - \frac{2\Phi}{c^2}\right) dl^2, \ee where $dl=|{\bf x}|$
denotes Euclidean arc length. The effect of the space-time
curvature on the propagation of light can be expressed in terms of
an effective index of refraction $n_{\rm eff}$ \cite{Schneider},
which is given by

\be
n_{\rm eff}=1 - \frac{2}{c^2} \Phi\,.
\ee
Thus, the effective speed of light in the field of a negative mass is
\be
v_{\rm eff} = c/n_{\rm eff} \approx c + \frac{2}{c} \Phi\,.
\ee

Because of the increase in the effective speed of light in the
gravitational field of a negative mass, light rays would arrive faster
than those following a similar path in vacuum. This leads to a very
interesting effect when compared with the propagation of a
light signal in the gravitational field of a positive mass. In that case,
light rays are delayed relative to propagation in vacuum---the
well known {\it Shapiro time delay}. In the case of a
negative mass lensing, this effect is replaced by a new one, which
we shall call {\it time gain}. We will describe this effect in
more detail in the following subsections.

Defining the deflection angle as the difference of the initial and
final ray direction
\be
\bbox{\alpha} : = \hat{\bf e}_{\rm in} -\hat{\bf e}_{\rm out},
\ee
where $\hat{\bf e}:=d {\bf x}/dl$ is the unit tangent vector of a ray
${\bf x}(l)$, we obtain the deflection angle as the integral along
the light path of the gradient of the gravitational potential
\be
\bbox{\alpha} = \frac{2}{c^2} \int \bbox{\nabla}_{\bot}\Phi dl\,\,,
\ee
where $\bbox{\nabla}_{\bot}\Phi$ denotes the projection of
$\bbox{\nabla}\Phi$ onto the plane orthogonal to the direction
$\hat{\bf e}$ of the ray. We find
\be
\bbox{\nabla}_{\bot}\Phi(b,z) = -\frac{G |M|{\bf b}}{(b^2
+z^2)^{3/2}}\,.
\ee
Eq.~7 then yields the deflection angle
\be
\bbox{\alpha} = -\frac{4 G |M| {\bf b}}{c^2 b^2}\,\,.
\ee

It is interesting to point out that in the case of the negative
mass lensing, the term `deflection' has its rightful meaning---the
light is deflected away from the mass, unlike in the positive mass
lensing, where it is bent towards the mass.

\subsection{Lensing geometry and lens equation}

\begin{figure}[ht!]
\centerline{\epsfig{figure=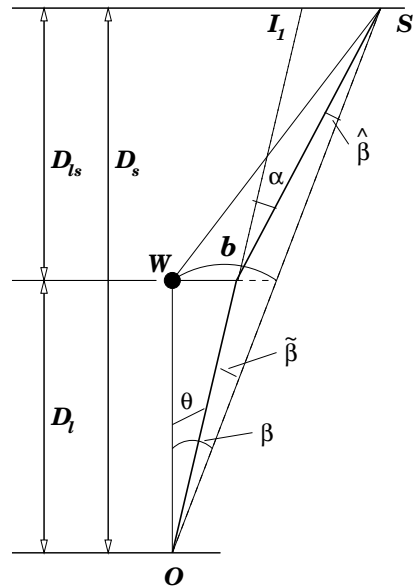,width=0.3\textwidth}}
\vspace*{.25cm}
\caption{Lensing geometry of a negative mass.
$O$ is the observer, $S$ is the source, $W$ is the negative mass
lens, $I_1$ is one of the images. $\beta$ is the angle between
the source and the lens---position of the source, $
\theta$ is the angle between the source and the image---position
of the image, and $\alpha$ is the deflection angle. $b$ is the impact
parameter and $D_l$, $D_s$ and $D_{ls}$ are angular diameter
distances. Other quantities are auxiliary.}
\label{fig:nlens}
\end{figure}

In Fig.~2 we present the lensing geometry for a point-like
negative mass. From this figure and the
definition of the deflection angle (Eq.~6), we can obtain the relation
between the positions of the source and the image. We only need to
relate the radial distance of the source and the image from the
center, since due to circular symmetry, the azimuthal angle $\phi$
is not affected by lensing. This gives
\be
(\bbox{\beta} - \bbox{\theta}) D_{\rm s} =-\bbox{\alpha} D_{\rm ls}
\ee
or
\be
\bbox{\beta} = \bbox{\theta} - \frac{D_{\rm {ls}}}{D_{\rm s}}
\bbox{\alpha}\,\,.
\ee

With the deflection angle (Eq.~9), we can write the
lens equation as
\be
\beta=\theta +\frac{4G|M|}{c^2 \xi}\frac{D_{ls}}{D_s}=\theta +
\frac{4G|M|}{c^2}\frac{D_{ls}}{D_s D_l} \frac{1}{\theta}\,\,.
\ee

\subsection{Einstein radius and the formation of images}

A natural angular scale in this problem is given by the quantity
\be
\theta^2_{\rm E}=\frac{4 G|M|}{c^2} \frac{D_{\rm ls}}{D_{\rm s}
D_{\rm l}}\,\,,
\ee
which is called the Einstein angle. In the
case of a positive point mass lens, this corresponds to the angle
at which the Einstein ring is formed, happening when source, lens
and observer are perfectly aligned. As we will see later in this
section, this does not happen if the mass of the lens is negative.
There are other differences as well. A typical angular separation
of images is of order $2\,\theta_{\rm E}$ for a positive mass lens.
Sources which are closer than about $\theta_{\rm E}$ to
the optical axis are significantly magnified, whereas sources which
are located well outside the Einstein ring are magnified very
little. All this is different with a negative mass lens, but nonetheless,
the Einstein angle remains a
useful scale for the description of the various regimes in the
present case and, therefore, we shall use the same nomenclature for
its definition.

The Einstein angle corresponds to the Einstein radius in the
linear scale (in the lens plane): \be R_{\rm E} = \theta_{\rm E}
D_{\rm l} = \sqrt{\frac{4 GM}{c^2} \frac{D_{\rm ls}D_{\rm
l}}{D_{\rm s}}}\,\,. \ee In terms of Einstein angle the lens
equation takes the form \be \beta=\theta+ \frac{\theta_{\rm
E}^2}{\theta}\,\,, \ee which can be solved to obtain two solutions
for the image position $\theta$: \be \theta_{1,2} = \frac{1}{2}
\left( \beta\pm \sqrt{\beta^2 - 4 \theta_{\rm E}^2}\right). \ee
Unlike in the lensing due to positive masses, we find that there
are three distinct regimes here and, thus, can classify the
lensing phenomenon as follows:

\begin{enumerate}

\item[I:] {$\beta < 2 \theta_{\rm E}$}~~There is no real solution for the
lens equation. It means that there are no images when the source is
inside twice the Einstein angle.

\item[II:] { $\beta > 2 \theta_{\rm E}$}~~There are two solutions,
corresponding to two images both on the same side of the lens and
between the source and the lens. One is always inside the Einstein
angle, the other is always outside it.

\item[III:] { $\beta=2\,\theta_{\rm E}$}~~This is a degenerate case,
$\theta_{1,2}=\theta_{\rm E}$; two images merge at the Einstein
angular radius, forming the {\it radial} arc (see \S~\ref{sec:mag}).
\end{enumerate}

We also obtain two important scales, one is the Einstein angle
($\theta_{\rm E}$)---the angular radius of the radial critical
curve, the other is twice the Einstein angle ($2\,\theta_{\rm
E}$)---the angular radius of the caustic. Thus, we have two
images, one is always inside the $\theta_{\rm E}$, one is always
outside; and as a source approaches the caustic ($2\,\theta_{\rm
E}$) from the positive side, two images coming closer and closer
together, and nearer and the critical curve, thereby brightening.
When the source crosses the caustic, the two images merge on the
critical curve ($\theta_{\rm E}$) and disappear.

\subsection{Time Gain and Time-Offset Function}

Following \cite{Narayan}, we define a scalar potential,
$\psi(\bbox{\theta})$, which is the appropriately scaled projected
Newtonian potential of the lens,
\be
\psi(\bbox{\theta}) =
\frac{D_{\rm ls}}{D_{\rm l} D_{\rm s}}\frac {2}{c^2}
\int{\Phi(D_{\rm l}\bbox{\theta},z) dz}\,\,.
\ee
For a negative point mass lens it is
\be
\psi(\bbox{\theta}) =
\frac{D_{\rm ls}}{D_{\rm l} D_{\rm s}} \frac{4 G |M|}{c^2}
\ln{|\theta|}\,\,.
\ee
The derivative of $\psi$ with respect to $\bbox{\theta}$ is
the deflection angle
\be
\bbox{\nabla}_{\theta}\psi
= D_l \bbox{\nabla}_b \psi = \frac{2}{c^2}\frac{D_{ls}}{D_s}
\int{\bbox{\nabla}_{\bot}\Phi dz} = \bbox{\alpha}\,\,.
\ee
Thus, the
deflection angle is the gradient of $\psi$---the deflection
potential
\be
\bbox{\alpha}(\bbox{\theta})= \bbox{\nabla}_\theta \psi\,\,.
\ee
>From this fact and the lens equation (11) we obtain
\be
(\bbox{\theta} -\bbox{\beta}) + \bbox{\nabla}_{\theta}
\psi(\bbox{\theta}) = 0 \,\,.
\ee
This
equation can be written as a gradient,
\be
\bbox{\nabla}_{\theta}\left[\frac{1}{2}(\bbox{\theta} -
\bbox{\beta})^2 + \psi(\bbox{\theta}) \right] = 0\,\,.
\ee
If we compare this equation
with that for the Fermat's principle \cite{Narayan}
\be
\bbox{\nabla}_{\theta}\,t(\bbox{\theta}) = 0\,\,,
\ee
we see that we can define the
time-offset function (opposite to time-delay function in the case
of positive mass lens) as
\be
t(\bbox{\theta}) = \frac{(1+z_{\rm
l})}{c} \frac{D_{\rm l} D_{\rm s}}{D_{\rm ls}} \left[\frac{1}{2}
(\bbox{\theta}-\bbox{\beta})^2 + \psi(\bbox{\theta}) \right] =
t_{\rm geom} + \tilde{t}_{\rm pot},
\ee
where $t_{\rm geom}$ is the geometrical
time delay due to the extra path length of the deflected light ray
relative to the unperturbed one. It remains the same as in the
positive case---increase of light-travel-time relative to an
unbent ray. The
coefficient in front of the square brackets ensures that the
quantity corresponds to the time offset as measured by the
observer. The second term $\tilde{t}_{\rm pot}$ is the time gain a
ray experiences as it traverses the deflection potential
$\psi(\bbox{\theta})$, with an extra factor $(1+z_l)$ for the
cosmological `redshifting'. Thus, cosmological geometrical time
delay is
\be
t_{\rm geom} = \frac{(1+z_{\rm l})}{c} \frac{D_{\rm
l} D_{\rm s}}{D_{\rm ls}} \frac{1}{2}(\bbox{\theta} - \bbox{\beta})^2\,\,,
\ee
and cosmological potential time gain is
\be
\tilde{t}_{\rm
pot}=\frac{(1+z_{\rm l})}{c} \frac{D_{\rm l} D_{\rm s}}{D_{\rm ls}}
\psi(\bbox{\theta})\,\,.
\ee

\begin{figure}
\leavevmode \epsfxsize=8cm \epsfysize=9.5cm \epsffile{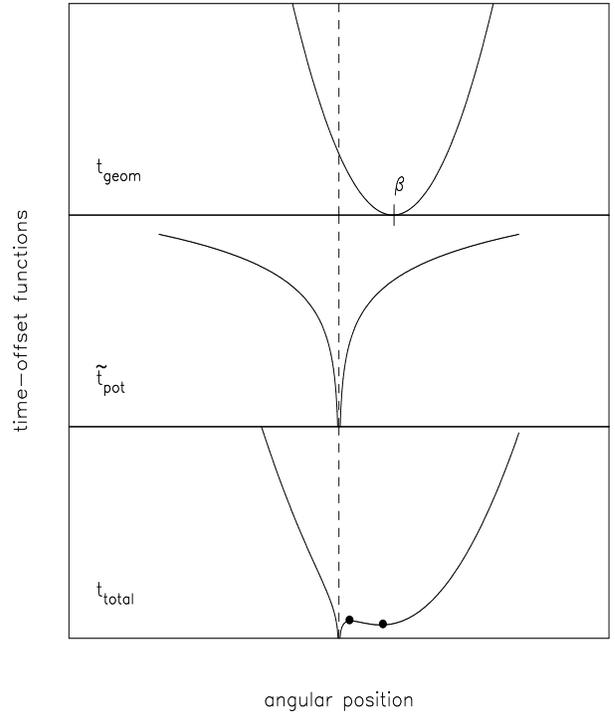}
\vspace{.25cm}\caption{Geometric time delay, gravitational time
gain and total time offset produced by a point negative mass lens
for a source that is slightly off the optical axis.}
\label{fig:time}
\end{figure}

In Fig.~\ref{fig:time} we show the time delay and time gain
functions. The top panel shows $\rm t_{\rm geom}$ for a slightly
offset source. The curve is a parabola centered on the position of
the source, $\beta$. The central panel displays $\rm \tilde t_{\rm pot}$
for a point negative mass lens. This curve is centered on the
lens. The bottom panel shows the total time-offset. Images are
located at the stationary points of $\rm t_{\rm total}$. Here we
see two extrema---maximum and minimum---on the same side (right)
from the optical axis (marked by dots).

We can find the time difference between the two images, $\theta_1$
and $\theta_2$, that is, if a source has intrinsic variability, it will appear
in the two images at an interval
\be
\D t_{12} = \frac{r_{\rm s}}{c} (1+z_{\rm l}) \left( \nu^{1/2} -
\nu^{-1/2} - \ln{\nu} \right),
\ee
where by $\nu$ we denoted the
ratio of absolute values of magnifications of images,
\be
\frac{\mu_{1}}{\mu_{2}} = \left[\frac{\sqrt{u^2-4}
+u}{\sqrt{u^2-4} - u} \right]^2\,\,,
\label{eq:ratio}
\ee
and $u$ is the scaled angular position of the source $u=\beta/\theta_{\rm E}$
and $r_{\rm s}$ is the Schwarzschild radius of the lens.

\subsection{Magnifications}\label{sec:mag}

Light deflection not only changes the direction but also the
cross-section of a bundle of rays. For an infinitesimally small
source, the ratio between the solid angles gives the flux
amplification due to lensing
\be
|\mu|=\frac{d\omega_i}{d\omega_s}\,\,.
\ee
For an
infinitesimal source at angular position $\bbox{\beta}$ and image
at angular position $\bbox{\theta}$, the relation between the
two solid angles is determined by the area distortion, given in
turn by the determinant of the Jacobian matrix $\cal A$ of the
lens mapping $\bbox{\theta} \mapsto \bbox{\beta}$
\be
{\cal A}
\equiv \frac{\partial \bbox{\beta}} {\partial \bbox{\theta}}\,\,.
\ee
For a point mass lens magnification is given by
\be
\mu^{-1} = \left|\frac{\beta}{\theta}\frac{d\beta}{d\theta}\right|\,\,.
\ee
The image is thus magnified or demagnified by a factor $|\mu|$. If
a source is mapped into several images, the total amplification is
given by the sum of the individual image magnifications. From the
lens equation (15), we find
\be
\frac{\beta}{\theta} =
\frac{\theta^2 + \theta_{\rm E}^2}{\theta^2}\,\,\,\,\,;\,\,\,
\frac{d\beta}{d\theta} = \frac{\theta^2 - \theta_{\rm
E}^2}{\theta^2}\,\,.
\ee
Thus,
\be \mu^{-1}_{1,2} = \left| 1 -
\frac{\theta_{\rm E}^4}{\theta_{1,2}^4}\right|\,\,,
\ee
and using $u$ from (28), we find the total magnification (Fig.~4,
bottom panel, continuous curve) as
\be
\mu_{\rm tot} = |\mu_1|
+|\mu_2| = \frac{u^2-2}{u \sqrt{u^2-4}}\,\,.
\ee

The tangential and radial critical curves follow from the
singularities in
\be
\mu_{\rm
tan}=\left|\frac{\beta}{\theta}\right|^{-1}
=\frac{\theta^2}{\theta^2+\theta_{\rm E}^2}
\ee
and
\be \mu_{\rm
rad} = \left|\frac{d\beta}{d\theta}\right|^{-1} =
\frac{\theta^2}{\theta^2-\theta_{\rm E}^2}\,\,.
\ee
$\mu_{\rm rad}$ diverges when $\theta=\theta_{\rm E}$---angular
radius of the radial critical curve. $\mu_{\rm tan}$
always remains finite, which means that there are no
tangential critical curves---no tangential arcs can be
formed by the negative point mass lens. In Fig.~4 we show the
magnification curves (radial, tangential and total) for both
positive (upper panel) and negative mass lenses (bottom panel).
The difference can be seen as follows---in the upper panel there
is no singularity in the radial curve (no radial arcs are formed
by the positive mass lens), whereas in the bottom panel
we see that the curve for the radial magnification experiences
a singularity.

\begin{figure}
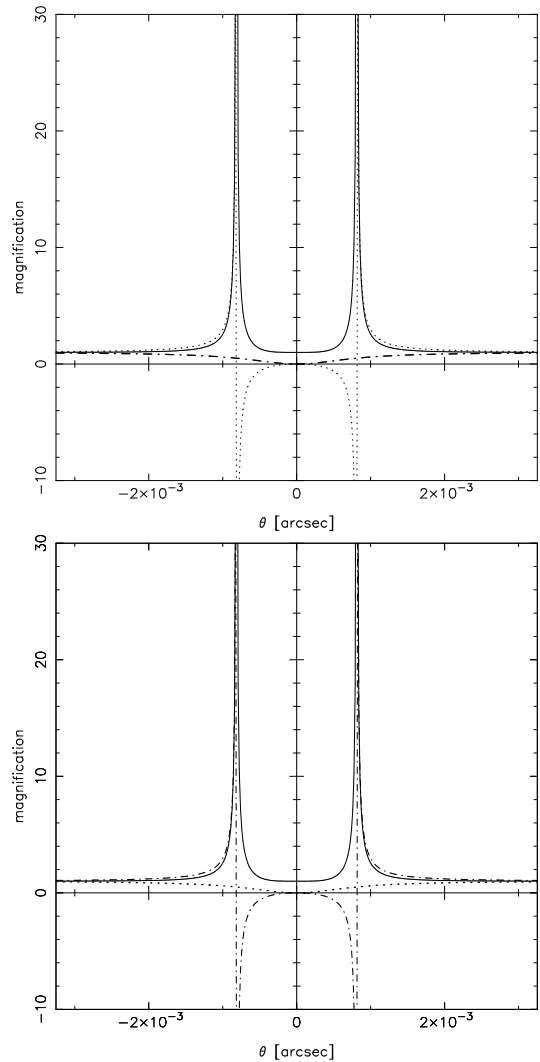

\begin{flushleft} \epsfxsize=7cm \epsfysize=7cm
\epsffile{mag_pos.ps}  \epsfxsize=7cm \epsfysize=7cm
\epsffile{mag_neg.ps} \end{flushleft} \vspace{.25cm}\caption{The
magnifications: tangential $\mu_{\rm tan}$ (dotted lines), radial
$\mu_{\rm rad}$ (dash-dotted lines), and total $\mu$ (continuous curves),
are plotted as functions of the image position $\theta$ for two cases;
in the upper panel for the positive mass, in the bottom panel
for the negative mass. The singularities of $\mu_{\rm tan}$
and $\mu_{\rm rad}$ give the positions of the tangential and
radial critical curves, respectively. In the upper panel the
singularity is in the tangential critical curve, in the bottom
panel, instead, in the radial critical curve. Here $|M|=1\,
M_{\odot}$, $D_{\rm s}=0.05$ Mpc and $D_{\rm l}=0.01$ Mpc.
Angles are in arcseconds.}
\end{figure}

\section{Microlensing}

\begin{figure}
\centerline{
\includegraphics[width=0.4\textwidth]{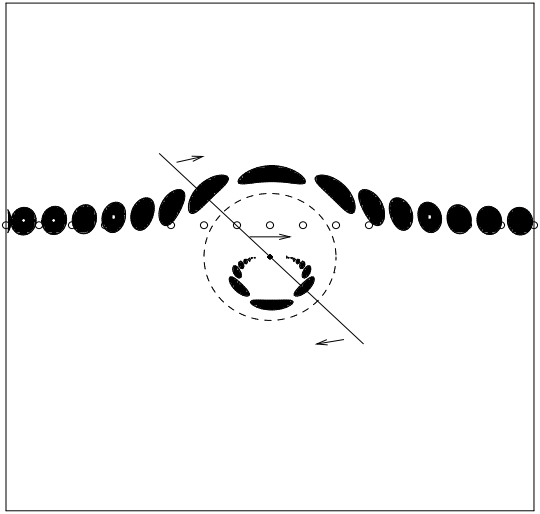}}
\vspace{.25cm} \caption{Schematic representation of the geometry
of the positive mass lensing due to the motion of the source,
lens and the observer (in this case we can consider only
the motion of the source in the plane perpendicular to the
optical axis). The lens is indicated with a dot at the
center of the Einstein ring, which is marked with a dashed line.
The positions of the source center are
shown with a series of small open circles. The locations and the
shapes of the two images are shown with a series of dark ellipses.
At any instant, the two images, the source and the lens are all on
a single line, as shown in the figure for one particular instant.
\label{fig:positive}}
\end{figure}

When the angular separation between the images $\D \theta$ \be \D
\theta= \sqrt{\beta^2 - 4\theta_{\rm E}^2} \,\, \ee is of the
order of milliarcsecs, we cannot resolve the two images with
existing telescopes and we can only observe the lensing effect
through their combined light intensity. This effect is called {\it
microlensing}. Both the lens and the source are moving with
respect to each other (as well as the observer). Thus, images
change their position and brightness. Of particular interest are
sudden changes in luminosity, which occur when a compact source
crosses a critical curve. For the positive mass lensing the
situation is quite simple (Fig.~\ref{fig:positive}) (for a review
on the positive mass microlensing and its applications, see
\cite{sazhin1}).

For a negative mass lens the situation is different. We define a
dimensionless minimum impact parameter $B_0$, expressed in terms
of the Einstein radius, as the shortest distance between the path
line of the source and the lens. For three different values of
$B_0$ we have three different lensing configurations shown in
Figs.~6,~7, and~8. Note the large difference in the shapes of the
images for these three regimes. In Fig.~9 we show the case of a
minimum impact parameter equal to zero, $B_0=0$, that is, the path
of the source passes through the lens.

\begin{figure}
\centerline{
\includegraphics[width=0.385\textwidth]{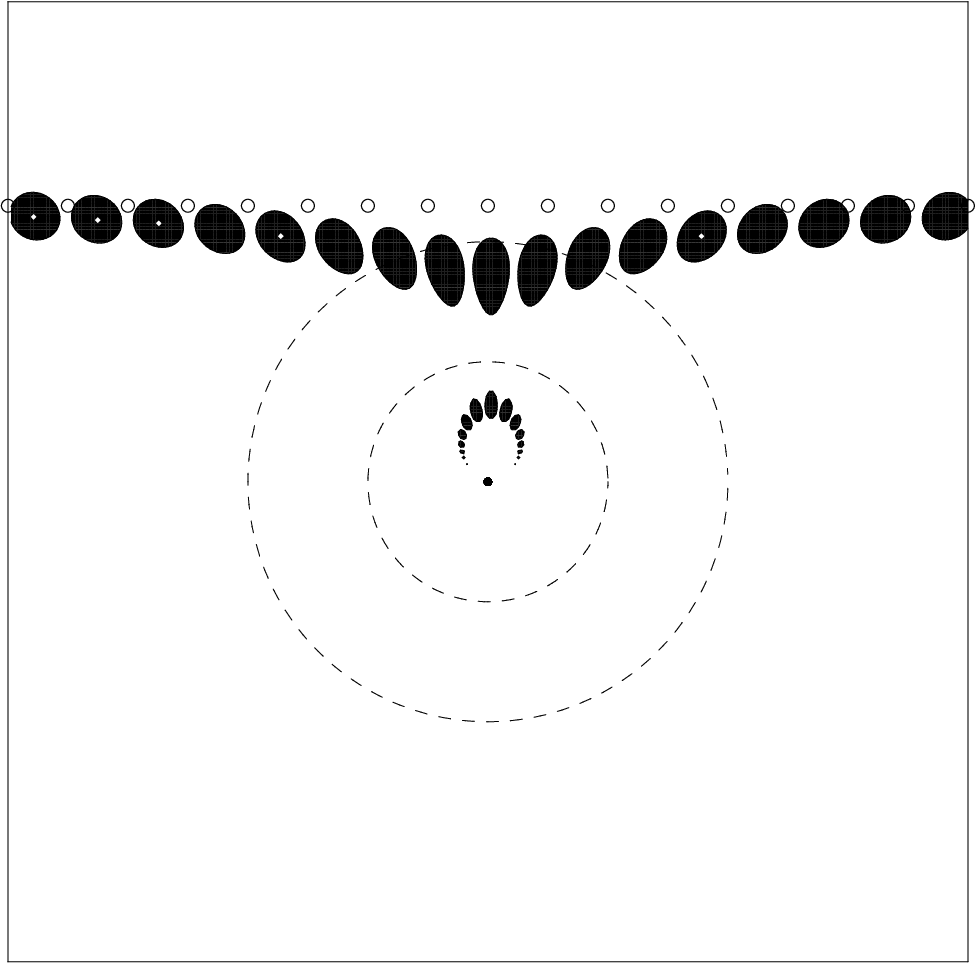}}
\vspace{.25cm} \caption{True motion of the source and apparent
motion of the images for $B_0 > 2$. The inner dashed circle is the
Einstein ring, the outer dashed circle is twice the Einstein ring.
The rest is as in Fig.~\ref{fig:positive}.}
\end{figure}

\begin{figure}
\centerline{
\includegraphics[width=0.385\textwidth]{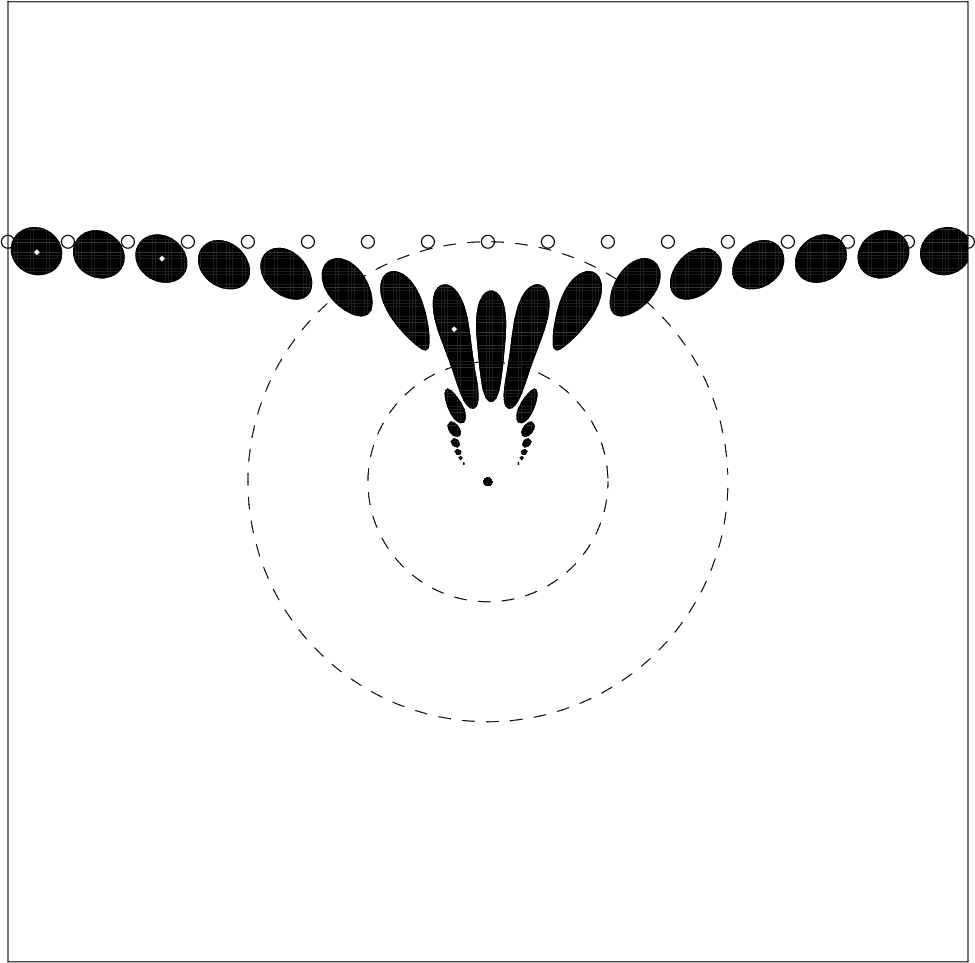}}
\vspace{.25cm} \caption{True motion of the source and apparent
motion of the images for $B_0= 2$. The inner dashed circle is the
Einstein ring, the outer dashed circle is twice the Einstein ring.
The rest is as in Fig.~\ref{fig:positive}.}
\end{figure}

\begin{figure}
\centerline{
\includegraphics[width=0.385\textwidth]{bmin_lt_2.ps}}
\vspace{.25cm} \caption{True motion of the source and apparent
motion of the images for $B_0 < 2$. The inner dashed circle is the
Einstein ring, the outer dashed circle is twice the Einstein ring.
The rest is as in Fig.~\ref{fig:positive}.}
\end{figure}

\begin{figure}
\centerline{
\includegraphics[width=0.385\textwidth]{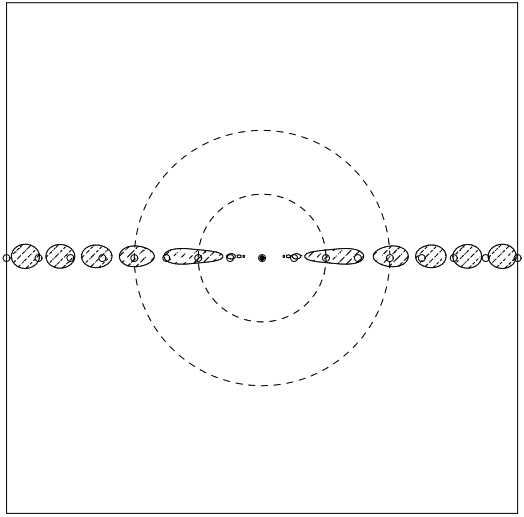}}
\vspace{.25cm} \caption{True motion of the source and apparent
motion of the images for $B_0=0$. The inner dashed circle is the
Einstein ring, the outer dashed circle is twice the Einstein ring.
Images here are shown with the shaded ellipses. The rest is as in
Fig.~\ref{fig:positive}.}
\end{figure}

It can be
assumed without the loss of generality that the observer and the lens are
motionless and the source moves in the plane perpendicular to the line
of sight (therefore, changing its position in the source plane).
We adopt the treatment given in \cite{Kayser1986} for the
velocity $V$, and consider effective transverse velocity of the source
relative to the critical curve (see Appendix A).
We define the time scale of the microlensing event as the time it
takes the source to move across the Einstein radius,
projected onto the source plane, $\xi_0=\theta_{\rm E} D_{\rm s}$,

\be 
t_{\rm v}=\frac{\xi_0}{V}\,\,. \label{eq:tscale} 
\ee 
Angle
$\beta$ changes with time according to 
\be 
\beta(t) =
\sqrt{\left(\frac{Vt}{D_{\rm s}}\right)^2 + \beta_0^2}\,\,. 
\ee
Here the moment $t=0$ corresponds to the smallest angular distance
$\beta_0$ between the lens and the source. Normalizing (39) to
$\theta_E$, 
\be 
u(t) = \sqrt{\left(\frac{Vt}{\theta_{\rm E} D_{\rm
s}}\right)^2 + \left(\frac{\beta_0}{\theta_{\rm E}}\right)^2}\,\,,
\ee 
where dimensionless impact parameter $u$ is defined in (28).
Including the time scale $t_v$ and defining \be B_0=
\frac{\beta_0}{\theta_{\rm E}}\,\,, \ee we obtain \be u(t)=
\sqrt{B_0^2 + \left(\frac{t}{t_v}\right)^2}\,\,. \label{eq:u(t)}
\ee Finally, the total amplification as a function of time is
given by \be A(t)=\frac{u(t)^2-2}{u(t)\sqrt{u(t)^2-4}}\,\,. \ee

Comparing this analysis with that of Cramer et al. \cite{cramer}, we must
note that they wrote the equation for the time dependent dimensionless
impact parameter as (cf. our Eq.~\ref{eq:u(t)})
$$
B(t) = B_0 \sqrt{1 + \left( \frac{t}{T_0} \right)^2}\,\,,
$$
and defined the time scale for the microlensing event as
the time it takes to cross the minimum
impact parameter (cf. our Eq.~\ref{eq:tscale})
$$
T_0=\frac{b_0}{V}\,\,,
$$
where $b_0$ is the minimum impact parameter and other variables
carry the same meanings as in our paper. While there is no mistake
in using such definitions, there is a definite disadvantage in
doing so. Using Eq.~10 of \cite{cramer} for $B(t)$ we cannot build
the light curve for the case of the minimum impact parameter
$B_0=0$. In this case their Eq.~8 diverges, although there is
nothing wrong with this value of $B_0$ (see our Figs.~9 and~10).
In the same way, their definition of a time scale does not give
much information on the light curves. With our definition
(Eq.~\ref{eq:tscale}) we can see in Fig.~10 that in the extreme
case of $B_0=0$ the separation between the half-events is exactly
$2\theta_{\rm E}$; it is always less than that with any other
value of $B_0$.

\begin{figure}
\begin{center} \epsfxsize=7cm \epsfysize=7cm \epsffile{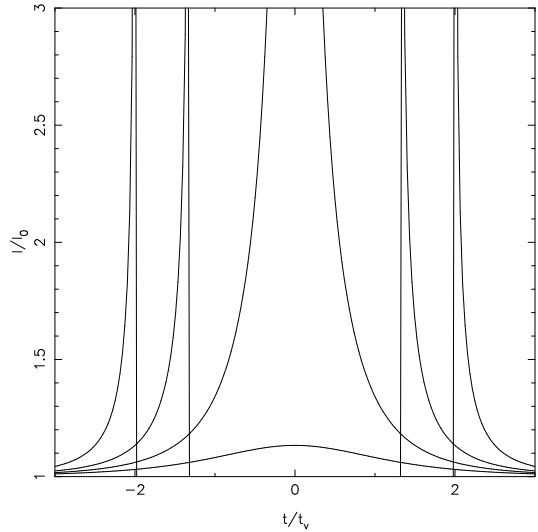}
\vspace{.25cm}\end{center}
\caption{Light curves for the negative mass lensing of a point source.
>From the center of the graph towards the corners the curves correspond
to $B_0=2.5$, $2.0$, $1.5$, $0.0$. The time scale here is $\xi_0$
divided by the effective transverse velocity of the source.
\label{fig:light}}
\end{figure}

In Fig.~\ref{fig:light} we show the light curves for the point source
for four source trajectories with different minimum impact parameters
$B_0$. As can be seen from the light curves, when the distance
from the point mass to the source trajectory is larger than
$2\theta_{\rm E}$, the light curve is identical to that of a
positive mass lens light curve. However, when the distance is less
than $2\theta_{\rm E}$ (or in other terms, $B_0 \le 2.0$), the light curve
shows significant differences. Such events are characterized by the {\it
asymmetrical} light curves, which occur when a compact source
crosses a critical curve. A very interesting, eclipse-like,
phenomenon occurs here; a zero intensity region
(disappearance of images) with an angular radius $\theta_0$
\be
\theta_0 = \sqrt{4\theta^2_{\rm E}- \beta_0^2}\,\,,
\ee
or in terms of normalized unit $\theta_{\rm E}$,
\be
\D = \sqrt{4 - B_0^2}\,\,.
\label{eq:Delta}
\ee

In the next Section we shall see how these features get affected
by the presence of an extended source.

\section{Extended Source}

In the previous sections we considered magnifications and light
curves for point sources. However, sources are
extended, and although their size may be small compared to the
relevant length scales of a lensing event, this extension
definitely has an impact on the light curves, as will be
demonstrated below. From variability arguments, the optical and
X-ray continuum emitting regions of quasars are assumed to be much
less than 1 pc~\cite{Schneider2}, whereas the broad-line emission
probably has a radius as small as $0.1$ pc~\cite{Osterbrock1976}.
The high energy gamma-spheres have a typical radius of 10$^{15}$
cm \cite{blan}. Hence, one has to consider a fairly broad range of
source sizes.

We define here the dimensionless source radius, $\tilde R$, as
\be
\tilde R = \frac{\rho}{\theta_{\rm E}} = \frac{R}{\xi_0}\,\,,
\ee
where $\rho$ and $R$ is the angular and the linear physical size
of the source, respectively, and $\xi_0$ is the the length unit
in the source plane (see Eq.~38).

\subsection{Comments on numerical method and simulations}

It is convenient to write the lens equation in the scaled scalar
form \be y = x + \frac{1}{x}\,\,, \label{eq:lens} \ee where we
normalized the coordinates to the Einstein angle: \footnote{Note,
that for the case in which $x$ and $y$ are expressed in length
units, we obtain a different normalization in each plane, which is
not always convenient.} 
\be x=\frac{\theta}{\theta_{\rm E}}
\,\,\,;\,\,\,\,\,\,\,y= \frac{\beta}{\theta_{\rm E}}\,\,.
\label{eq:eqscaled} 
\ee 
The lens equation can be solved
analytically for any source position. The amplification factor,
and thus the total amplification, can be readily calculated for
point sources. However, as we are interested in extended sources,
this amplification has to be integrated over the source
(Eq.~\ref{eq:total}), and furthermore, as we want to build the
light curves, the total amplification for an extended source has
to be calculated for many source positions. The amplification
${\cal A}$ of an extended source with surface brightness profile
$I({\bf y})$ is given by 
\be 
{\cal A} = \frac{\int d^2y I({\bf y})
{\cal A}_0({\bf y})} {\int d^2y I({\bf y})}\,\,, 
\label{eq:total}
\ee 
where ${\cal A}_0({\bf y})$ is the amplification of a point
source at position ${\bf y}$. We have used the numerical method
first described in \cite{Schramm1987}. We cover the lens plane
with a uniform grid. Each pixel on this grid is mapped, using
Eq.~\ref{eq:lens}, into the source plane. The step width ($5000
\times 5000$) is chosen according the desired accuracy (i.e. the
observable brightness). For a given source position
($y_{01},y_{02}$) we calculate the squared deviation function
(SDF) \be S^2=(y_{10} - y_1(x_1,x_2))^2 +
(y_{20}-y_2(x_1,x_2))^2\,\,. \label{eq:sdf} \ee The solutions of
the lens equation (Eq.~\ref{eq:eqscaled}) are given by the zeroes
of the SDF. Besides, Eq.~\ref{eq:sdf} describes circles with radii
$S$ around ($y_{10},y_{20}$) in the source plane. Thus, the lines
$S=$ const are just the image shapes of a source with radius $S$,
which we plot using standard plotting software. Therefore, image
points where SDF has value $S^2$ correspond to those points of the
circular source which are at a distance $S$ from the center. The
surface brightness is preserved along the ray and  if $I(R_0)=I_0$
for the source, then the same intensity is given to those pixels
where SDF$=R_0^2$. In this way an intensity profile is created in
the image plane and integrating over it we can obtain the total
intensity of an image. Thus, we obtain the approximate value of
the total magnification by estimating the total intensity of all
the images and dividing it by that of the unlensed source,
according to the corresponding brightness profile of the source
(see Appendix B). For a source with the constant surface
brightness the luminosity of the images is proportional to the
area enclosed by the line $S=$ const. And the total magnification
is obtained by estimating the total area of all images and
dividing it by that of the unlensed source. For calculations of
light curves we used a circular source which is displaced along a
straight line in the source plane with steps equal to 0.01 of the
Einstein angular radius.

\subsection{Results}

In Figs.~11 and~12 we show four projected source and image
positions, critical curves/caustics in the lens/source plane and
representative light curves for different normalized source sizes.
The sources are taken to be circular disks with constant surface
brightness. In order to get absolute source radii and real light
curves we need the value of $\theta_{\rm E}$, the normalized
angular unit, the distance to the source, as well as the velocity
$V$ of the source relative to the critical curves in the source
plane (see Appendix A). We have used $M=M_{\odot}$, $H_0=100$ km
s$^{-1}$ Mpc$^{-1}$ and a standard cosmological model with zero
cosmological constant. Here and in all subsequent simulations the
redshift of the source is $z_{\rm s}=0.5$ and the redshift of the
lens is $z_{\rm l}=0.1$.

We display two cases for two different impact parameters. It must be
noted here that the minimum impact parameter $B_0$ defines now the
shortest distance between the line of path of the center of the
source and the lens. For each one of $B_0$, the dimensionless
radius of the source $\tilde R$ increases from 0.01 to 2.0 in
normalized units of $\theta_{\rm E}$. The shape of produced images changes
notably with the increase of the source size, as can be seen in
the bottom right panel of Figs.~11 and~12. At the same time the
smaller the source the greater the magnification, since when
the source radius is greater than the Einstein radius of the lens,
the exterior parts, which are amplified, compete with the interior
ones, which are demagnified.

It can be noted that, despite the noise in some of the simulated
light curves, the sharp peaks which occur when the source is
crossing the critical line are well defined even for the smallest
source. Note that all infinities are replaced now by finite
amplifications, and that the curves are softened; all these
effects being generated by the finite size of the source. Indeed,
while the impressive drop to zero in the light curve is
maintained, the divergence to infinity, that happens for a point
source, is very much reduced. Note, that in cases of a large size
of the source the magnification is very small. If we would like to
see bigger enhancement than that, we should consider sources of
smaller sizes, approaching the point source situation (cf.
Fig.~11, upper left plot).

It is also interesting to note here that for the impact parameter 
$B_0=2.0$ the light curve of a small extended source, though 
approaching the point source pattern (Fig.~10), still differs 
considerably from it (Fig.~12, upper left plot).

\begin{figure}
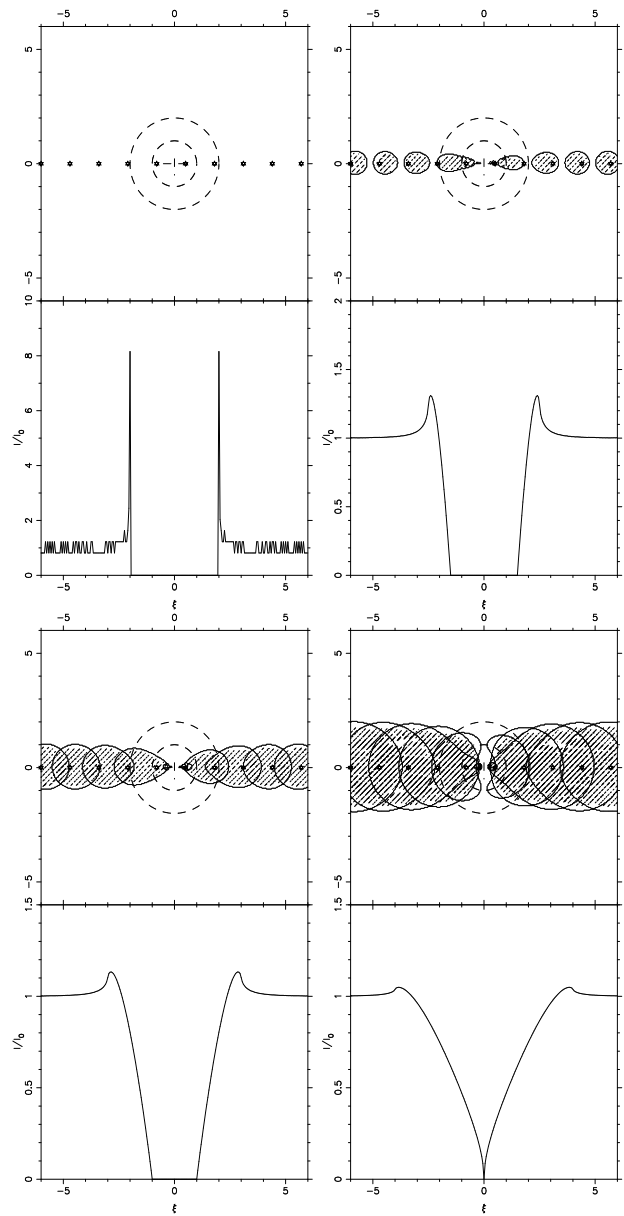

\begin{flushleft} \epsfxsize=4cm \epsfysize=8cm
\epsffile{ext0_0.ps}  \epsfxsize=4cm \epsfysize=8cm
\epsffile{ext0_4.ps}  \epsfxsize=4cm \epsfysize=8cm
\epsffile{ext0_5.ps}  \epsfxsize=4cm \epsfysize=8cm
\epsffile{ext0_7.ps}
\end{flushleft} \vspace{.25cm}
\caption{Four sets of lens-source configurations
(\emph{upper panels}) and corresponding amplification
as a function of source's center position (\emph{bottom panels})
are shown for four different values of the dimensionless
source radius $\tilde R $ (0.01, 0.5, 1.0, 2.0, in normalized units,
$\theta_{\rm E}$). Each of the four upper panels display
the time dependent position of the source's center, the shapes
of images (shaded ellipses) and critical curves (dashed circles).
The series of open small circles show the path of the source center.
The lens is marked by the central cross. Minimum impact parameter
$B_0=0$. By replacing $\xi$ with $\xi \theta_{\rm E} D_{\rm s}
V^{-1} = \xi t_{\rm v}$ we get corresponding time depending light curve.}
\end{figure}

\begin{figure}
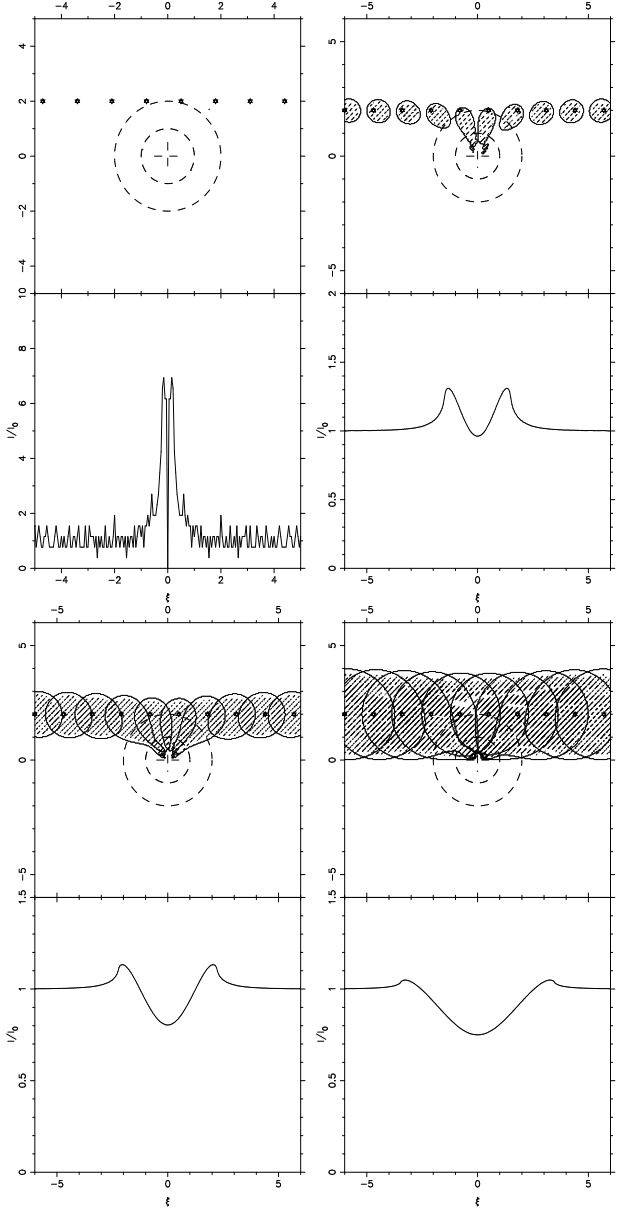

\begin{flushleft} \epsfxsize=4cm \epsfysize=8cm
\epsffile{ext2_0.ps}  \epsfxsize=4cm \epsfysize=8cm
\epsffile{ext2_4.ps}  \epsfxsize=4cm \epsfysize=8cm
\epsffile{ext2_5.ps}  \epsfxsize=4cm \epsfysize=8cm
\epsffile{ext2_7.ps}
\end{flushleft} \vspace{.25cm}
\caption{Same as in Fig.~11, but for minimum impact parameter
$B_0=2.0$.}
\end{figure}

In Figs.~13 and~14 we show the images of an extended source with a
Gaussian brightness distribution for two effective dimensionless
source radii (Appendix B), $\tilde R_{\rm S}$, of 3.0 and 1.0
(Fig.~13, frames a--e and Fig.~14, frames a--e, respectively),
together with the corresponding light curves (Fig.~13, frame f and
Fig.~14, frame f, respectively). Here the source path passes
through the lens ($B_0=0$), which lies exactly in the center of
each frame. In Fig.~13 the source's extent in the lens plane is
greater than the Einstein radius of the lens. Annotated wedges 
provide colour scale for the images. We notice there that
there is an eclipse-like phenomenon, occurring most notably when
most of the source is near or exactly behind the lens. This is
consistent with the light curve (frame f), where there is a
de-magnification. For the source with radius smaller than the
double Einstein radius of the lens (Fig.~14), the low intensity
region is replaced by the zero intensity region; the source
completely disappears from the view (frame c).

In Fig.~15 we display the images of the source with the
exponential brightness profile (see Appendix B) and the
corresponding light curve (frame f). The effective radius of the
source is 1.5. The impact parameter here is $B_0=2.0$; the lensing
regime corresponds to the one schematically depicted in Fig.~7. We
see how shapes of the images change, becoming elongated and
forming the radial arc (frames c and d).

In order to compare a constant surface brightness source with more
realistic distributions, we simulate images configurations and
calculate light curves for two different assumed profiles with
radial symmetry (see Appendix B).

\begin{figure}
\centerline{\epsfig{figure=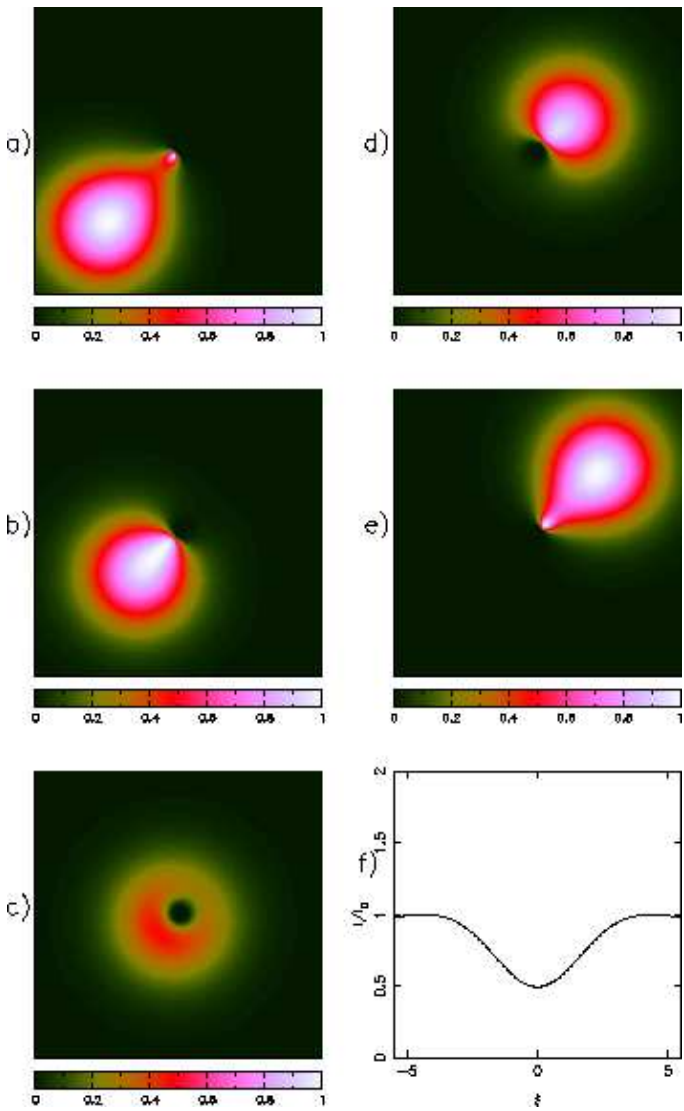,width=0.5\textwidth}}
\vspace{.25cm}
\caption{Image configurations (frames a to e) and a corresponding
light curve (f) for a Gaussian source with effective radius
$\tilde R_{\rm S}=3.0$, in units of the Einstein angle. The source
is moving from the lower left corner (frame a) to the right
upper corner (frame e), passing through the lens ($B_0=0$).
The lens is in the center of each frame. Size of each frame
is $5 \times 5$, in the normalized units. Wedges to each frame
provide the colour scale for the images. Note the eclipse-like
phenomenon, consistent with the incomplete
demagnification showed in the light curve (frame f).
\label{fig:gauss1}}
\end{figure}

\begin{figure}
\centerline{\epsfig{figure=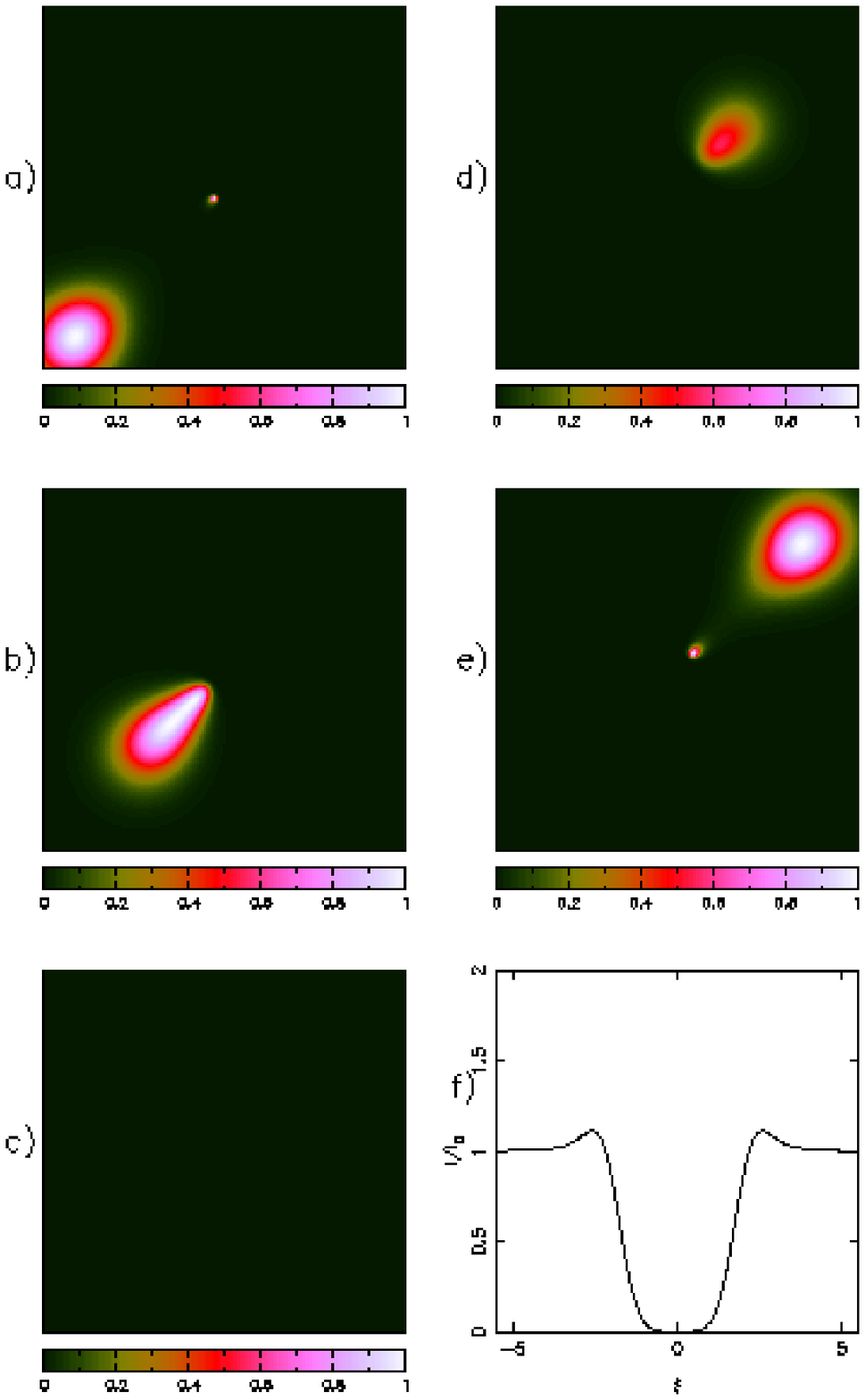,width=0.5\textwidth}}
\vspace{.25cm}
\caption{Image configurations (frames a to e) and a corresponding
light curve (f) for a Gaussian source with effective radius
$\tilde R_{\rm S}=1.0$, in units of the Einstein angle. The source
is moving from the lower left corner (frame a) to the right
upper corner (frame e), passing through the lens ($B_0=0$).
The lens is in the center of each frame. Size of each frame
is $3 \times 3$, in the normalized units. Wedges to each frame
provide the colour scale for the images. Note the complete
disappearence of the source when it is inside the double Einstein
radius of the lens (frame c), corresponding to the drop
of magnification to zero in the light curve (frame f).
\label{fig:gauss2}}
\end{figure}

\begin{figure}
\centerline{\epsfig{figure=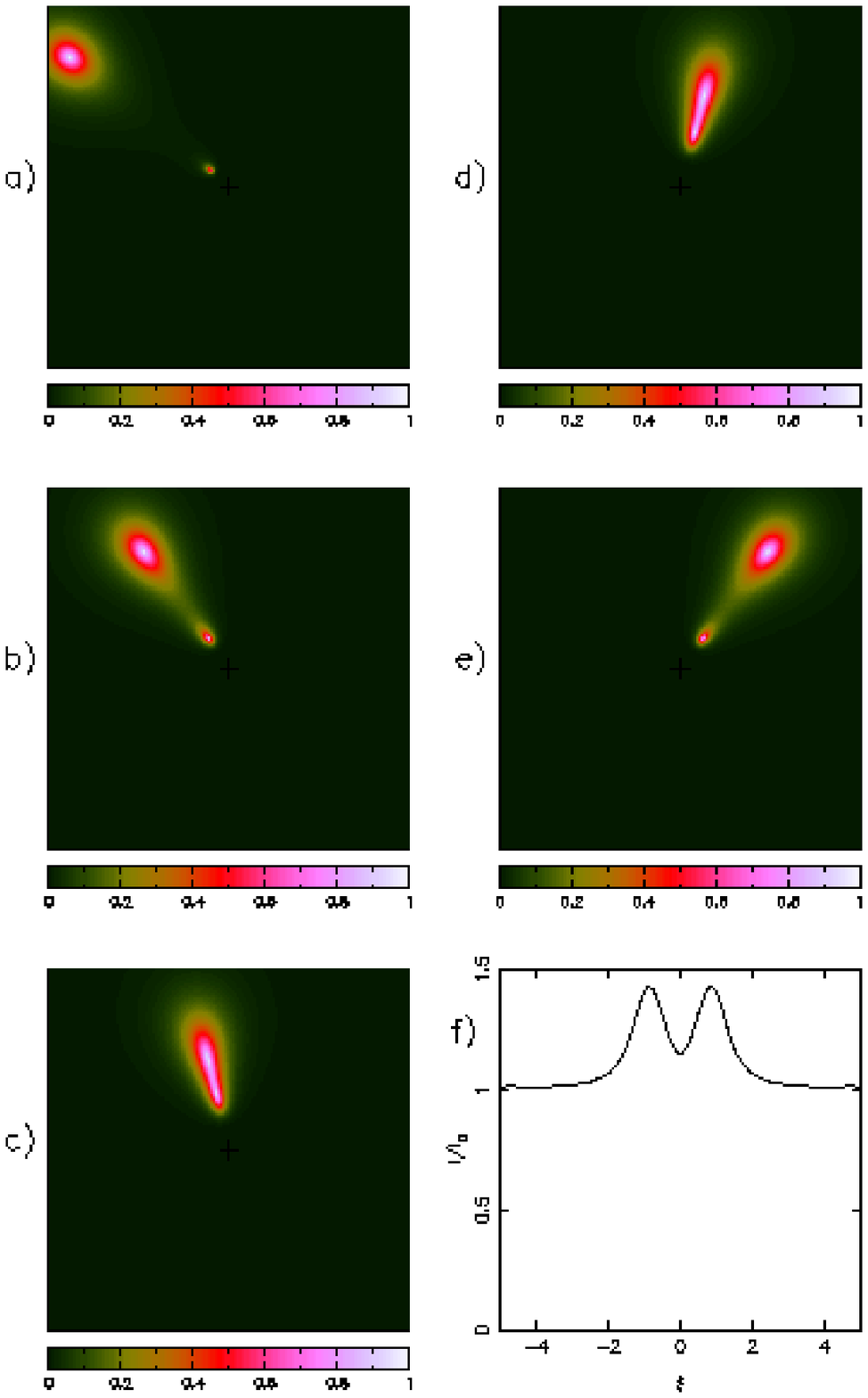,width=0.5\textwidth}}
\vspace{.25cm}
\caption{Image configurations (frames a to e) and a
corresponding light curve (f) for a source
with the exponential brightness distribution and effective radius
$\tilde R_{\rm S}=1.0$, in units of the Einstein angle. The source
is moving from the upper left corner (frame a)
to the upper right corner (frame e) with the impact parameter
$B_0=2.0$. The lens is in the center of each frame. Size of each
frame is $2.5 \times 2.5$, in the normalized units. Wedges to each frame
provide the colour scale for the images.
\label{fig:exponent}}
\end{figure}

\begin{figure}[ht!]
\begin{center}
\leavevmode
\vspace*{.25cm}
\epsfxsize=7cm
\epsffile{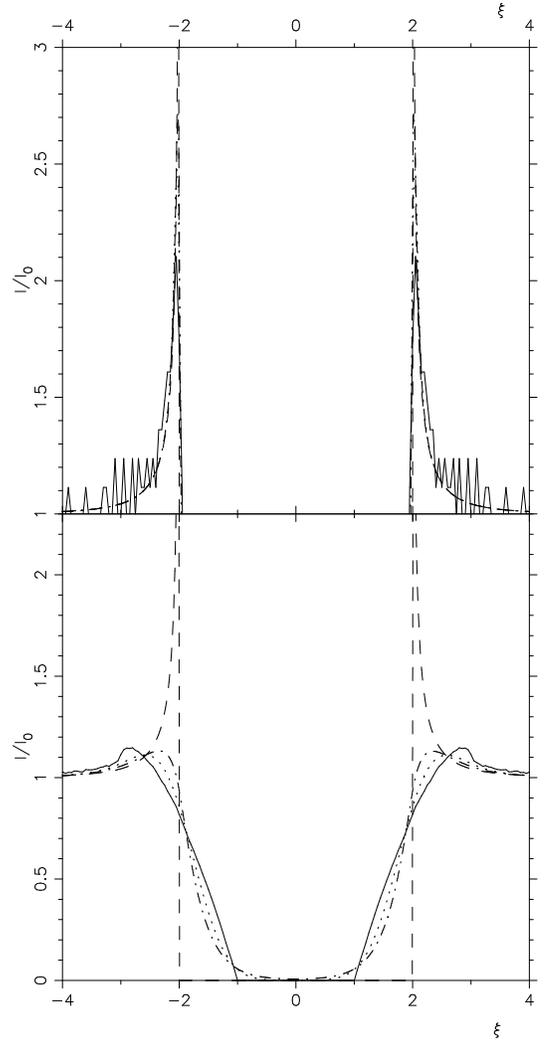}
\end{center}
\caption{Light curves for the point mass source (dashed line),
source with constant surface brightness (solid),
source with Gaussian brightness distribution (dash-dotted)
and exponential brightness distribution (dotted) for
two different effective dimensionless source radii, $0.1$
(\emph{upper panel}) and $1.0$ (\emph{bottom panel}).
\label{fig:lcurves}}
\end{figure}

In Fig.~\ref{fig:lcurves} we compare light curves for three
different radially symmetric source profiles, uniform, Gaussian
and exponential, for two dimensionless source radii $\tilde R =
\tilde R_{\rm S}^{\rm gauss} = \tilde R_{\rm S}^{\rm expon} = 0.1$
and $\tilde R=\tilde R_{\rm S}^{\rm gauss} = \tilde R_{\rm S}^{\rm
expon} = 1.0$. As a reference curve we show the light curve of the
point source. All curves are made for the impact parameter
$B_0=0$. We can see the larger noise in the uniform source curve,
since the source with uniform brightness has extremely sharp edge,
whereas Gaussian and exponential sources are extremely smooth.
Though we considered the sources with the same effective radius,
we can see from the plot that for a small source size, the maximum
magnification is reached by the  source with exponential profile
(upper panel), which is explained by the fact that this profile
has a more narrow central peak than the Gaussian.

For the larger source, this behaviour smoothens, though we still
can see large differences in the light curves (bottom panel),
where the uniform source experiences darkening, while sources with
other profiles only undergo demagnification.

\section{Concluding remarks}
In this paper we have explored the consequences, regarding
gravitational microlensing, of the existence of matter violating
the energy conditions. We have also quantitatively analyzed, using
numerical simulations, the influence of a finite size of the
source on the gravitational lensing negative-mass event. We have
thus enhanced and completed previous works, where the focus was
put on the point source light curves and no discussion was given 
concerning the shapes of images, actual simulations of microlensing 
events, time gain function, and other features presented here. 
Figs.~4,~6--9 and~11--16 comprise our new results: a useful 
comparison arena where to test observationally the possible 
existence of wormholes or any other kind of negative mass 
compact objects.

The next step would be to test these predictions using archival,
current, and forthcoming observational microlensing experiments.
The only search done up to now included the BATSE database of
gamma-ray bursts, but there is still much unexplored territory in
the gravitational microlensing archives. We suggest to adapt
the alert systems of these experiments in order to include the
possible effects of negative masses as well. This, perhaps, would
not imply a burdensome work, but there could be a whole new world
of discoveries.

\acknowledgments
MS is supported by a ICCR scholarship (Indo-Russian Exchange
programme) and acknowledges the hospitality of IUCAA, Pune. We
would like to deeply thank Tarun Deep Saini for his invaluable
help with the programming and Prof. Daksh Lohiya and Prof. M.~V.
Sazhin for useful discussions. This work has also been
supported by CONICET (PIP 0430/98, GER), ANPCT (PICT 98 No.
03-04881, GER) and Fundaci\'on Antorchas (through separate
grants to DFT and GER). D.F.T. especially acknowledges
I. Andruchow for her continuing support.

\appendix

\section{Velocities and time scales}

Let the source have a transverse velocity ${\bf v}_{\rm s}$
measured in the source plane, the lens a transverse velocity ${\bf
v}_{\rm l}$ measured in the lens plane, and the observer a
transverse velocity ${\bf v}_{\rm obs}$ measured in the observer
plane. The effective transverse velocity of the source relative to
the critical curves with time measured by the observer is \be {\bf
V} = \frac{1}{1+z_{\rm s}}{\bf v}_{\rm s} - \frac{1}{1+z_{\rm l}}
\frac{D_{\rm s}}{D_{\rm l}}{\bf v}_{\rm l} + \frac{1}{1+z_{\rm l}}
\frac{D_{\rm ls}}{D_{\rm l}}{\bf v}_{\rm obs}\,\,, \ee where this
effective velocity is such that for a stationary observer and
lens, the position of the source will change in time according to
$\delta \bbox{\xi}={\bf V} \Delta t$.

We basically have two time scales of interest here. The first one 
is the typical rise time to a peak in the amplification. We can 
estimate that it corresponds to a displacement of $\Delta y 
\sim \tilde R$ of the source across a critical line; the 
corresponding time scale $\tau_1$ is 
\be 
\tau_1 = t_{\rm v} \tilde R\,\,, 
\ee 
where $t_{\rm v}$ is given by (\ref{eq:tscale}), or in terms 
of the physical source size $R=\xi_0 \tilde R$, 
\be 
\tau_1=\frac{R}{V}\,\,, 
\ee
where effective transverse velocity of the source $V$ is given by
(A1). The second time scale of interest is the time between two
peaks $\tau_2$. For a point source we can estimate it as 
\be
\tau_2=t_{\rm v} \Delta  \,\,, 
\ee 
where $\Delta$ is given in Eq.~(\ref{eq:Delta}). For a source 
with radius $\tilde R$ and impact parameter $B_0=0.0$ it can 
be shown to be 
\be 
\tau_2=t_{\rm
v} \left(\Delta + \tilde R\right)  \,\,. 
\ee
In Table I we list for different values of source radii the time scales
$\tau_1$ and $\tau_2$, and the time delay between two images $\Delta t_{12}$ 
for the point source. The value of $V$ we estimate to be $V=5000$ km s$^{-1}$.

\noindent

\begin{table}
\caption{Time scales for several source radii. $\tilde R$ is the 
dimensionless source radius, in units of Einstein angle, 
$|M|=1.0\,M_{\odot}$, redshift of the lens is
$z_{\rm l}=0.1$, redshift of the source is $z_{\rm s}=0.5$, 
$\xi_0=5.42\times10^{11}$ km is the normalized length unit 
in the source plane. The time scales correspond
to apparent source velocity (see Eq.~A1) $V=5000$ km s$^{-1}$.}
\vskip 0.05in
\begin{tabular}{l|c|c|c|r}
\emph{$\tilde R$} & \emph{$R$} (pc) & \emph{$\Delta
t_{12}$}\tn{u=4.0 (definition in Eq.~\ref{eq:ratio})} (sec) &
\emph{$\tau_1$} (yr) & \emph{$\tau_2$}\tn{$B_0=0.0$} (yr)  \\
\hline
        &                     &                               &          &           \\
$0.0$   & point source        &      $2.0\times 10^{-4}$      &    --    & $6.78$    \\
$0.01$  & $1.07\times10^{-4}$ &             --                & $0.03$   & $6.81$    \\
$0.1$   & $1.07\times10^{-3}$ &              --               & $0.34$   & $7.11$    \\
$1.0$   & $1.07\times10^{-2}$ &              --               & $3.38$   & $10.16$   \\
$2.0$   & $0.3\times10^{-1}$  &               --              & $6.75$   & $13.6$    \\
\end{tabular}
\vskip 0.05in
\end{table}

\section{Extended sources brightness profiles}

\subsection{Source of uniform brightness}

For a circular source of radius $R$ and uniform brightness,
equation (49) transforms into 
\be {\cal A} = \frac{\int{d^2 y
I({\bf y}) {\cal A}_0({\bf y})}} {\pi R^2} \,\,.
\ee 

\subsection{Source with Gaussian brightness distribution}

For a Gaussian source we have 
\be 
I(r)=I_0 e^{-r^2/r_0^2}\,\,, 
\ee
where we normalized the profile such that the maximum value of $I$
equals unity. We define the radius containing $90\%$ of all the
luminosity as the effective radius of the source, $R_{\rm S}$. To
find the relation between $R_{\rm S}$ and $r_0$, we write the
total luminosity as 
\be 
L(\infty) = \int_{0}^{\infty}
e^{-{r^2}/{r_0^2}} 2 \pi r\,dr= \pi r_0^2 \,\,,
\ee
\be
L(R) = \int_{0}^{R} e^{- {r^2}/{r_0^2}} 2 \pi r\,dr= \pi r_0^2
\left[1 - e^{- {R^2}/{r_0^2}} \right]\,\,, 
\ee 
then 
\be \frac{L(R_{\rm
S})}{L(\infty)} = 0.9 = \left[ 1- e^{-{R_{\rm S}^2}/{r_0^2}} \right]\,\,. 
\ee 
Thus, effective radius relates to the parameter $r_0$
as 
\be 
\frac{R_{\rm S}}{r_0} = \sqrt{\ln{10}}\,\,. 
\ee

\vskip 0.2in

\subsection{Source with exponential brightness distribution}

We have 
\be 
I(r) = I_0 e^{-r/r_0}\,\,. 
\ee 
In the same manner as above,
$R_{\rm S}$ is defined as radius, containing $90\%$ of total
luminosity. In the same way as above, total luminosity
\be 
L(\infty) = \int_{0}^{\infty} e^{- {r}/{r_0}} 2 \pi r\,dr= 2
\pi r_0^2 \,\,,
\ee 
then 
\be 
L(R) = \int_{0}^{R} e^{- {r}/{r_0}} 2 \pi
r\,dr = 2 \pi \left[r_0^2 - \left(R \,r_0 + r_0^2\right)e^{-
{R}/{r_0}} \right]\,\,, 
\ee 
and 
\be 
\frac{L(R_{\rm S})}{L(\infty)} = 0.9 \,\,. 
\ee 
From where we find that effective radius relates to the
parameter $r_0$ as 
\be 
e^{- {R_{\rm S}}/{r_0}}\left(1+
\frac{R_{\rm S}}{r_0}\right) = 0.1 \,\,. 
\ee 
Solution to this equation gives $R_{\rm S}/r_0 \approx~ 3.89$. 
This profile is also normalized such that the maximum value 
of $I$ equals unity.

\end{document}